\documentclass[conference]{IEEEtran}
\IEEEoverridecommandlockouts

\usepackage{cite}
\usepackage{float}
\usepackage{booktabs}
\usepackage{multirow}
\usepackage{tabularx}
\usepackage{makecell}
\usepackage{multicol}
\usepackage{bm}
\usepackage{amsmath,amssymb,amsfonts}
\usepackage{algorithmic}
\usepackage{graphicx}
\usepackage{textcomp}
\usepackage{xcolor}
\usepackage[hidelinks]{hyperref}
\hypersetup{breaklinks=true}

\def\BibTeX{{\rm B\kern-.05em{\sc i\kern-.025em b}\kern-.08em
    T\kern-.1667em\lower.7ex\hbox{E}\kern-.125emX}}

\usepackage{fancyhdr}
\newcommand{\IEEEcopyrightcover}{%
  \thispagestyle{empty}%
  \onecolumn
  \begin{flushleft}
    \vspace*{0.25cm}
    {\Huge IEEE Copyright Notice}\\[1cm]
    {\normalsize
    © 2025 IEEE. Personal use of this material is permitted. Permission from IEEE must be obtained for all other uses, \\
    in any current or future media, including reprinting/republishing this material for advertising or promotional purposes, \\
    creating new collective works, for resale or redistribution to servers or lists, or reuse of any copyrighted component of \\
    this work in other works.\\[14pt]

    Accepted for publication in the \textit{2025 IEEE 102nd Vehicular Technology Conference (VTC2025-Fall)}.\\
    The Version of Record will appear in IEEE Xplore.\\[8pt]

    }
    \vfill
  \end{flushleft}
  \newpage
  \twocolumn
  \setcounter{page}{1} % restart numbering so main paper starts at page 1
}

\begin{document}

% Cover page (not counted, no header/footer)
\IEEEcopyrightcover

% Title and authors
\title{Exploring LR-FHSS Modulation for Enhanced IoT Connectivity: A Measurement Campaign}

\author{
    \IEEEauthorblockN{Alexis Delplace}
    \IEEEauthorblockA{
        \textit{Université Paris-Saclay}\\
        Palaiseau, France \\
        alexis.delplace@universite-paris-saclay.fr
        \vspace{0.2cm}
    }
\and
    \IEEEauthorblockN{Samer Lahoud}
    \IEEEauthorblockA{
        \textit{Faculty of Computer Science} \\
        \textit{Dalhousie University}\\
        Halifax, Canada \\
        sml@dal.ca
    }
\and
    \IEEEauthorblockN{Kinda Khawam}
    \IEEEauthorblockA{
        \textit{DAVID Lab} \\
        \textit{Université de Versailles Saint-Quentin}\\
        Versailles, France \\
        kinda.khawam@uvsq.fr
    }
}

\maketitle
\thispagestyle{fancy}

\begin{abstract}
This paper presents the first comprehensive real-world measurement campaign comparing LR-FHSS and LoRa modulations within LoRaWAN networks in urban environments. Conducted in Halifax, Canada, the campaign used a LoRaWAN platform capable of operating both modulations in the FCC-regulated US915 band. Real-world measurements are crucial for capturing the effects of urban topology and signal propagation challenges, which are difficult to fully replicate in simulations. Results show that LR-FHSS can achieve up to a 20\,\% improvement in Packet Reception Rate (PRR) over traditional LoRa in dense urban areas. Additionally, the study investigated path loss and Received Signal Strength Indicator (RSSI), finding that LR-FHSS achieved a minimum RSSI of $-138~\mathrm{dBm}$ compared to LoRa's $-120~\mathrm{dBm}$. The findings demonstrate that the introduction of LR-FHSS enhances communication robustness and reliability under regulatory limitations and suggest promising applications in LoRaWAN networks.
\end{abstract}


\section{Introduction}

LoRaWAN (Long Range Wide Area Network) is a wireless communication technology designed for Low-Power Wide-Area Networks (LPWAN), specifically targeting Internet of Things (IoT) applications. The physical layer of LoRaWAN relies on LoRa (Long Range) modulation, which has become a global standard, widely adopted across various IoT applications. However, LoRa modulation has certain limitations. Since LoRaWAN operates in unlicensed frequency bands, its usage is subject to strict regulatory constraints. For example, in FCC-regulated regions (North America), the \emph{dwell time} limitation (see Section III) reduces the maximum packet length that can be transmitted using LoRaWAN \cite{FCC_90-233}. 

To address these limitations, a new modulation technique called LR-FHSS (Long Range Frequency Hopping Spread Spectrum) was introduced in recent versions of LoRaWAN \cite{LoRaWANRegionalParameters2021}. LR-FHSS combines the advantages of spread spectrum modulation with frequency hopping, effectively reducing interference and enhancing communication robustness. This new modulation also improves spectral efficiency and allows for longer transmission times while still complying with regulations.

Initial studies on LR-FHSS have explored its integration into LoRaWAN, focusing on frequency usage, data segmentation, and frame structure, supported by performance simulations \cite{boquet2020lr}. Empirical studies by Semtech have provided the first real-world evidence of its performance through interference and range tests \cite{semtech2022an1200}. Additionally, a prototype using real-world packet traces, supported by a custom LR-FHSS software demodulator, has offered detailed insights into LR-FHSS's operation in field conditions \cite{traces2020understanding}. Moreover, an open-source discrete-event simulator has been developed to model LR-FHSS networks, enabling the study of network dynamics and scalability \cite{ana2024lr}. Recent work has aimed to improve the performance of LR-FHSS through advanced techniques, although these efforts rely largely on assumptions and simulations rather than practical measurements \cite{LR-FHSS_aloha}.

Despite these advancements, the practical experimentation and application of LR-FHSS remain limited, with only a few real-world measurements, such as those by Semtech, which are based on data collected from only three geographic locations. Real-world measurements are essential for validating theoretical models and simulations, and provide an accurate assessment of LR-FHSS performance.

In contrast to prior studies, this work presents the first extensive real-world measurement campaign that directly compares LR-FHSS and LoRa within LoRaWAN networks under urban conditions. This campaign provides valuable data on the performance of both modulations in FCC-regulated environments, offering insights into their behavior across different data rates. Our results show that LR-FHSS, particularly with the DR5 (Data Rate) configuration, enhances packet reception rates by up to 20\,\% over traditional LoRa in dense urban areas, where interference is a notable challenge. Furthermore, the measurement campaign reveals significant variations in path loss and RSSI behavior, highlighting the limitations of theoretical models in complex urban environments. To facilitate further research and verification, the complete dataset collected during this campaign is publicly available \cite{datasetgithub}.

The paper is structured as follows: Section II details LoRa modulation, Section III describes LR-FHSS principles, Section IV outlines the measurement methodology, Section V presents results and analysis, and Section VI concludes with key findings and LR-FHSS applications in LoRaWAN networks.

\section{LoRa Overview}

\begin{figure}[H]
    \centerline{\includegraphics[width=0.495\textwidth]{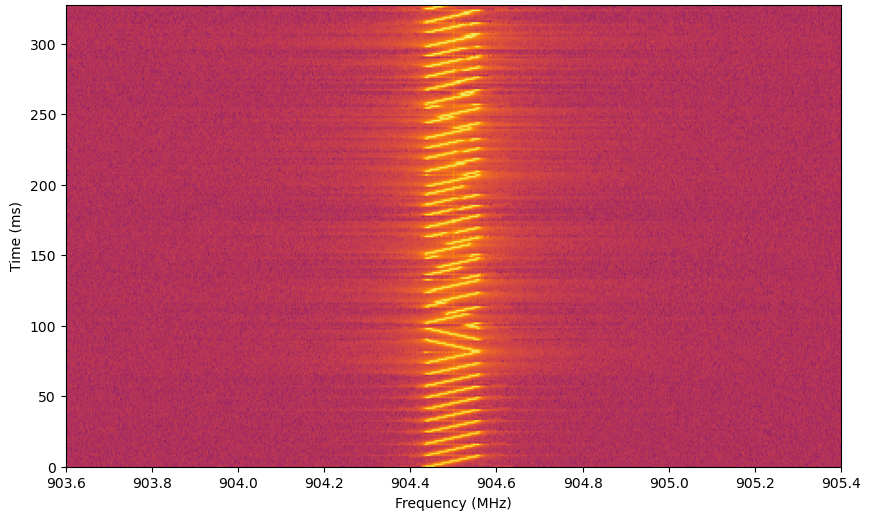}}
    \caption{Waterfall plot of LoRa transmission using DR0 (SF10, $125~\mathrm{kHz}$ bandwidth)}
    \label{fig_lora_modulation}
\end{figure}

LoRa (Long Range) is a spectrum-spreading modulation technique specifically designed for LPWANs. It is optimized for long-range communication with low power consumption, making it ideal for IoT applications. One of the key advantages of LoRa is its ability to demodulate signals below the noise floor, with a signal-to-noise ratio (SNR) as low as $-20~\mathrm{dB}$. This capability enables LoRa to achieve communication ranges that exceed $15~\mathrm{km}$ in rural areas and several kilometers in urban environments, where obstacles can significantly affect signal propagation \cite{el-chall:2019qy}. LoRa's robustness and range have led to its widespread adoption in scenarios requiring reliable connectivity over large geographic areas.

LoRa employs Chirp Spread Spectrum (CSS) modulation, a technique in which each symbol is represented by a chirp signal, a waveform whose instantaneous frequency increases (up-chirp) or decreases (down-chirp) linearly over the symbol period. This approach allows LoRa signals to sweep across the available bandwidth, providing resilience against interference and the Doppler effect. Fig.~\ref{fig_lora_modulation} illustrates this modulation, with time on the vertical axis and frequency on the horizontal axis. The bright diagonal lines represent the chirps, where the frequency shifts continuously over time and occupies a wide frequency range. This wideband nature ensures that even if part of the band is affected by noise or interference, the chirp can still be decoded from the unaffected portions. As a result, LoRa maintains reliable communication even in challenging environments, a robustness that will be assessed in the Measurement Campaign section of this paper.

The key parameters characterizing LoRa modulation include bandwidth (BW) and spreading factor (SF). LoRaWAN networks can operate at bandwidths of $125~\mathrm{kHz}$, $250~\mathrm{kHz}$, or $500~\mathrm{kHz}$, with wider bandwidths allowing faster data transmission but increasing susceptibility to interference.

The SF determines the duration of the chirp, balancing data rate and the overall link budget, which encompasses resistance to interference, noise, and path loss. In LoRaWAN, SF values range from 7 to 12. For example, with a bandwidth of $125~\mathrm{kHz}$, a lower SF such as SF7 provides a higher data rate (around $5.5~\mathrm{kbps}$), but results in a shorter communication range and lower robustness to interference. Conversely, a higher SF such as SF12 increases the link budget, enhancing resistance to interference and noise, and significantly extending the communication range, but reduces the data rate to approximately $250~\mathrm{bps}$. In LoRaWAN, the combination of BW and SF defines the Data Rate (DR) configuration, which specifies the modulation settings used for communication and can be adjusted to optimize network performance depending on deployment requirements.

\section{LR-FHSS Overview}

\begin{figure}[htbp]
    \centerline{\includegraphics[width=0.495\textwidth]{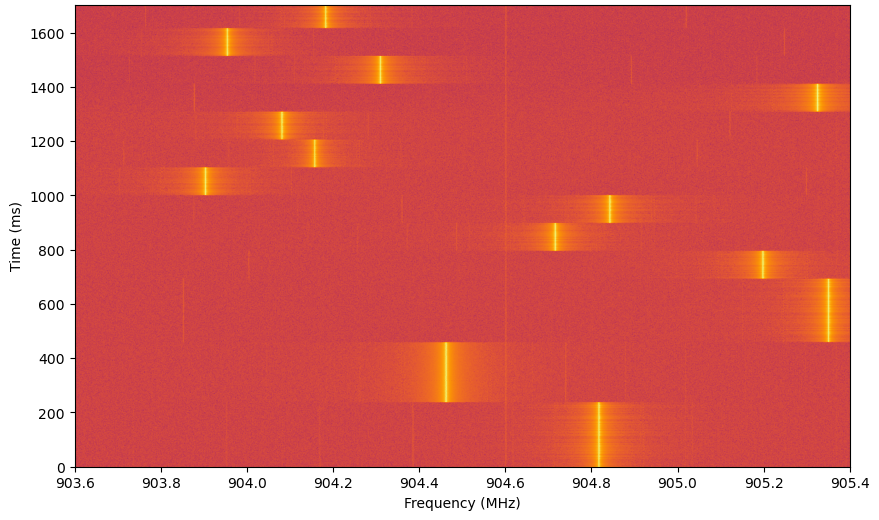}}
    \caption{Waterfall plot of LR-FHSS transmission using DR5 (CR~1/3, $1.523~\mathrm{MHz}$ bandwidth)}
    \label{fig_lrfss_modu}
\end{figure}

Recently, LR-FHSS (Long Range Frequency Hopping Spread Spectrum) was proposed as an enhancement to LoRaWAN networks to improve resilience and range in dense environments. LR-FHSS is a spread spectrum modulation technique that rapidly switches the carrier frequency across a wide range of available frequencies \cite{LoRaWANRegionalParameters2021}. By distributing the signal across multiple sub-channels, the frequency-hopping mechanism enhances resilience to interference by reducing the impact of localized disruptions. The hopping mechanism also adds a layer of security, making it more difficult for unintended receivers to intercept and piece together the transmission without knowing the hop sequence. According to \cite{semtech2022an1200}, LR-FHSS has a similar link budget to LoRa SF12BW125. However, in the US915 band, spectrum regulations limit LoRaWAN to a maximum spreading factor of SF10. Under these restrictions, LR-FHSS provides a better link budget and transmission range than LoRa SF10BW125.

In LoRaWAN, the operating channel width (OCW) for LR-FHSS is $1.523~\mathrm{MHz}$, subdivided into $488~\mathrm{Hz}$ sub-channels, referred to as Occupied Bandwidth (OBW). Packets are fragmented, with each fragment transmitted over a different sub-channel. The modulation used within these channels is Gaussian Minimum Shift Keying (GMSK). The sub-channels are organized in grids, with each grid comprising 60 sub-channels of $488~\mathrm{Hz}$, spaced $25.4~\mathrm{kHz}$ apart. These details are specific to the US915 band, where a frequency hop must have a minimum separation of $25~\mathrm{kHz}$ to meet regulatory requirements \cite{FCC_90-233}. In other LoRaWAN regions, the frequency hopping definition and grid organization may differ to comply with local regulations.

The waterfall plot in Fig.~\ref{fig_lrfss_modu} illustrates the LR-FHSS modulation, where the signal frequency hops across sub-channels over time. The plot depicts a LoRaWAN packet transmitted using DR5 (LR-FHSS), with the vertical axis representing time and the horizontal axis indicating frequency. The bright lines correspond to individual LR-FHSS fragments that constitute the packet. The first three fragments, each $233~\mathrm{ms}$ in duration, consist of repeated header transmissions, while the remaining fragments represent the payload, with durations of $102~\mathrm{ms}$ each, except for the last fragment, which is shorter due to the packet data not filling the entire fragment duration.

LR-FHSS offers several advantages over LoRa modulation in LoRaWAN networks, achieving similar or greater communication range and improved packet reception rates in noisy RF environments, thanks to its superior link budget compared to SF10BW125 in the US915 band. LR-FHSS also overcomes the single packet length limitation imposed on LoRa by the $400~\mathrm{ms}$ dwell time. In the US915 band, LoRaWAN packets using LoRa modulation are limited to an $11~\mathrm{bytes}$ payload in the best-case scenario (SF10BW125), whereas LR-FHSS packets are not subject to a theoretical length limit imposed by regulations. In practice, the maximum packet length is constrained by hardware capabilities, currently up to $125~\mathrm{bytes}$ \cite{boquet2020lr}. This capability makes LR-FHSS particularly advantageous for applications requiring larger data transmissions or more complex messages. Furthermore, LR-FHSS supports higher network capacity, which is crucial in scenarios with a low density of gateways compared to end-devices, such as satellite communications, where a single satellite can cover vast areas and serve many devices. Ongoing tests, such as those conducted by EchoStar Mobile, are evaluating the use of LR-FHSS for satellite LoRaWAN communications \cite{EchoStarMobile2023}, highlighting the industry's interest in its potential applications in constrained environments.

\section{Measurement Campaign}

\subsection{Measurement Setup}

The integration of LR-FHSS into LoRaWAN is relatively recent, with Semtech introducing the technology in 2020. Currently, no commercially available products utilize LR-FHSS. For end-devices, which are IoT devices responsible for transmitting data in a LoRaWAN network, Semtech has developed a kit for developers to test and become familiar with LR-FHSS. Some example code is provided for these tests, but support remains limited. For gateways, certain models can demodulate LR-FHSS packets, but this requires a firmware update that is not publicly available and must be obtained directly from Semtech.

\begin{figure}[htbp]
    \centerline{\includegraphics[width=0.495\textwidth]{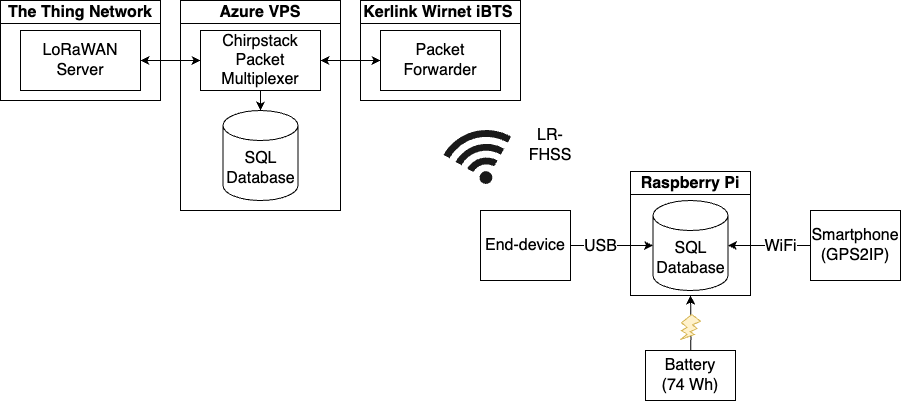}}
    \caption{Schematic Overview of the LR-FHSS Testbed Setup}
    \label{fig_plateforme}
\end{figure}

An end-to-end LoRaWAN platform was deployed for this measurement campaign, capable of operating both LoRa and LR-FHSS modulations. The platform consists of two main components for transmission and reception, respectively. Fig.~\ref{fig_plateforme} provides a schematic overview of the entire testbed setup.

The transmission component utilizes a Semtech LR1110MB1LCKS \cite{lr1110mb1lcks} board, which is a shield featuring Semtech's LoRa Edge LR1110, configured for the $915~\mathrm{MHz}$ frequency band used in North America. The LR1110 chip is fully compatible with LR-FHSS and supports intra-packet frequency hopping, unlike the previous generation SX1272/76 chips that lack this capability. The shield mounts on a STMicroelectronics NUCLEO-L476RG microcontroller and is equipped with a $3~\mathrm{dBi}$ antenna. This combination transmits LoRaWAN packets using both LoRa and LR-FHSS modulations. The firmware running on the microcontroller is based on the publicly available code provided by Semtech \cite{SWL2001}.

The microcontroller connects to a Raspberry Pi, enabling logging of all transmitted packet information into an SQL database. For each transmitted packet, the following information is recorded: date, time, frequency, data rate, spreading factor, coding rate, bandwidth, transmission power, and payload.

A smartphone connects to the Raspberry Pi via Wi-Fi and sends GPS coordinates, which the Pi aggregates with each transmitted packet's information and stores in the database.

The reception component consists of a Kerlink Wirnet iBTS gateway with a sensitivity of $-141~\mathrm{dBm}$ when used with LoRa SF12 \cite{wirnet_ibts}. The gateway is installed on the roof of the Goldberg building at Dalhousie University in Halifax, Canada, positioned $60~\mathrm{m}$ above sea level and $12~\mathrm{m}$ above ground level. The gateway firmware was updated to enable the reception of LR-FHSS packets, as this capability is not available with the default firmware. The updated firmware is provided by Semtech and is not publicly available.

A packet multiplexer logs information received by the gateway, sending one copy of the UDP packets to The Things Network (TTN) and another to a database, recording reception parameters for each packet.

\subsection{Measurement Protocol}

The measurement campaign aimed to collect a comprehensive dataset to enable a robust comparison between LR-FHSS and LoRa. The campaign was conducted in Halifax, Nova Scotia, Canada. Halifax's coastal geography presents a mix of dense urban environments in the downtown core, characterized by medium-rise buildings and narrow streets, and suburban areas with lower building density. Additionally, the city's hills and proximity to the Atlantic Ocean introduce varying terrain heights and potential signal reflections. These factors created a realistic testbed for evaluating LR-FHSS and LoRa performance under diverse propagation conditions, allowing for the collection of data representative of real-world performance in both challenging urban and open suburban environments.

\begin{figure}[htbp]
    \centerline{\includegraphics[width=0.325\textwidth]{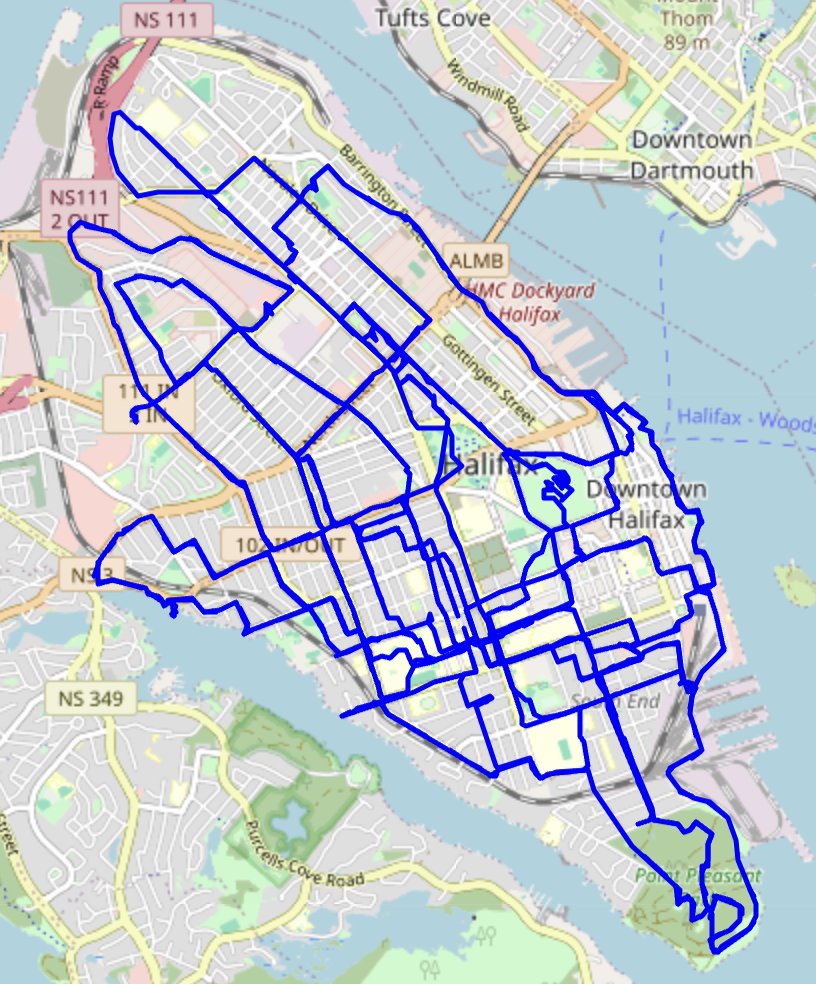}}
    \caption{Map of Halifax illustrating the extensive coverage of the measurement campaign, with blue trails representing the paths taken, covering more than 30\% of the city's streets}
    \label{fig_coverage}
\end{figure}

The campaign spanned approximately one month. This allowed for extensive coverage across both urban and suburban areas. All measurements were taken on foot at an average speed of $4~\mathrm{km/h}$, covering a total of $85~\mathrm{km}$.

The end-device was mounted on the end of a pole, positioned at a height of approximately $2~\mathrm{m}$. The device transmitted a message every $10~\mathrm{s}$, alternating between DR0, DR5, and DR6, with their configurations detailed in Table~\ref{tab:data_rate_configurations}. Each message consisted of a $4~\mathrm{bytes}$ payload transmitted at the maximum allowed power of $22~\mathrm{dBm}$, resulting in the collection of 7,435 data points. DR0 (LoRa SF10, $125~\mathrm{kHz}$) was selected for comparison with LR-FHSS as it offers the maximum link budget and is the largest spreading factor permitted for uplink in the US915 band, making it optimal for range-focused evaluations. Using a lower spreading factor would result in a lower link budget and reduced communication range, and therefore would not serve the purpose of testing the maximum achievable range.

For each data point, the following information was recorded: date, frequency, data rate, payload, latitude, longitude, altitude, distance to the gateway, RSSI, and SNR (for LoRa), as well as frequency drift (for LR-FHSS) and frequency offset.

\begin{table}[htbp]
    \centering
    \caption{Data Rate Configurations \cite{lora2021}}
    \begin{tabular}{>{\centering\arraybackslash}m{1cm} >{\centering\arraybackslash}m{1.5cm} >{\centering\arraybackslash}m{1.25cm} >{\centering\arraybackslash}m{1cm} >{\centering\arraybackslash}m{1cm}}
        \toprule
        \textbf{Data Rate} & \textbf{Modulation} & \textbf{Bandwidth} & \textbf{Coding Rate} & \makecell{\textbf{Bitrate} \\ \textbf{($\mathrm{bit/s}$)}} \\
        \midrule
        DR0 & LoRa SF10 & $125~\mathrm{kHz}$ & 4/5 & 980 \\
        DR5 & LR-FHSS & $1.523~\mathrm{MHz}$ & 1/3 & 162 \\
        DR6 & LR-FHSS & $1.523~\mathrm{MHz}$ & 2/3 & 325 \\
        \bottomrule
    \end{tabular}
    \label{tab:data_rate_configurations}
\end{table}

\section{Results}

\subsection{Performance Comparison: LoRa vs. LR-FHSS}

Our measurement campaign provides valuable insights into the performance of LR-FHSS and LoRa across various data rates in urban environments, where performance in this study specifically refers to Packet Reception Rate (PRR). The data includes measurements for DR0, DR5, and DR6 from our campaign in an urban setting, as well as from Semtech's study in a suburban environment in Neuchâtel, Switzerland~\cite{semtech2022an1200}. Semtech's data points are included for comparison, as they represent the only available real-world packet transmission data.

Notably, our measurements were conducted with a transmission power of $22~\mathrm{dBm}$ at a frequency of $915~\mathrm{MHz}$, whereas Semtech's study used $14~\mathrm{dBm}$ at $868~\mathrm{MHz}$. Despite using the EU frequency band, Semtech employed US parameters for data rate configurations. Their study collected data from three fixed locations with varying elevations at distances of $1.5~\mathrm{km}$, $3~\mathrm{km}$, and $6.3~\mathrm{km}$ from the gateway. In contrast, our campaign covered over $30\,\%$ of Halifax's streets, providing a diverse range of data points across the city. This comprehensive approach enables a more robust analysis, as our results at each distance are based on multiple measurements taken at locations around that distance from the gateway, while Semtech's results rely on isolated points.

\begin{figure}[htbp]
    \centerline{\includegraphics[width=0.495\textwidth]{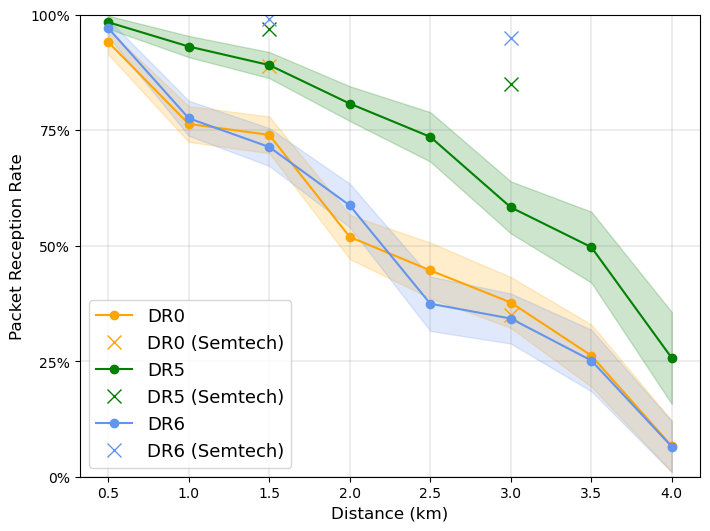}}
    \caption{Comparison of packet reception rate as a function of distance between the transceiver and the gateway for DR0 (LoRa), DR5 (LR-FHSS), and DR6 (LR-FHSS), with 95\% confidence intervals computed using the bootstrap technique.}
    \label{fig_PRR}
\end{figure}

The results highlight similar PRR between DR0 (LoRa) and DR6 (LR-FHSS), while DR5 (LR-FHSS) achieves statistically significant improvements. For instance, at $1.5~\mathrm{km}$, DR0 and DR6 show $75\,\%$ PRR, whereas DR5 achieves $85\,\%$. Using the bootstrap technique, we computed a $95\,\%$ confidence interval for the PRR improvement of DR5 over DR0, which is 0.1918 with a confidence interval of [0.1663, 0.2167]. These results suggest that LR-FHSS modulation (used in DR5) offers better performance, particularly in dense urban environments where interference is a major challenge. However, DR0 has a significantly shorter time on air ($330~\mathrm{ms}$) compared to DR5 ($1\,713~\mathrm{ms}$), resulting in lower energy consumption for DR0.

The maximum transmission distances recorded were $3.88~\mathrm{km}$ for LoRa and $3.97~\mathrm{km}$ for LR-FHSS. This suggests that LR-FHSS improves the signal's ability to traverse complex urban structures rather than extending the absolute communication range.

The performance gap between DR5 and DR6 aligns with their coding rate configurations (Table~\ref{tab:data_rate_configurations}). While LR-FHSS's frequency hopping provides inherent interference resistance and regulatory compliance for extended frames, this mechanism alone does not guarantee performance gains. Interestingly, DR6 fails to outperform DR0 in our measurements, contrasting with Semtech's results where DR6 performed better. The combination of frequency hopping and the stronger coding rate in DR5 enhances error correction, making it more resilient in challenging urban environments. While adopting an even more robust coding rate could further improve PRR, it would also increase packet duration and energy consumption.

The higher PRR achieved by LR-FHSS implies that fewer gateways may be needed for full city coverage compared to LoRa, while maintaining the same quality of service. Reducing the number of required gateways could lower infrastructure costs and simplify network management, making LR-FHSS an attractive option for IoT deployments in urban areas with space or budget constraints.

\subsection{Path Loss Analysis}

\begin{figure}[htbp]
    \centerline{\includegraphics[width=0.495\textwidth]{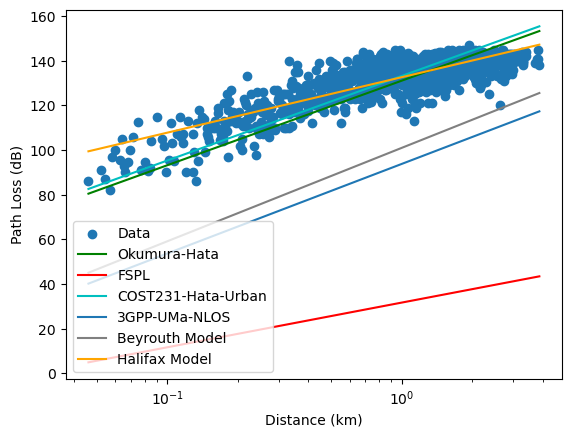}}
    \caption{Comparison of measured path loss from our urban field measurements with theoretical models (Okumura-Hata, Free Space Path Loss, COST231-Hata-Urban, 3GPP UMa NLOS), an empirical model from LoRa measurements in Beirut, and an empirical model derived from our data.}
    \label{fig_pathloss}
\end{figure}

Our measurement campaign reveals notable differences in path loss behavior, highlighting the limitations of theoretical models in predicting signal propagation in urban environments. Fig.~\ref{fig_pathloss} compares the measured path loss from our campaign with various models, including Okumura-Hata~\cite{okumura1968field}, Free Space Path Loss (FSPL), COST231-Hata-Urban~\cite{cost2311999digital}, 3GPP UMa NLOS~\cite{3gpp2010utra}, and an empirical model from LoRa measurements in Beirut, Lebanon~\cite{el-chall:2019qy}. We also derived a simple empirical model from our data using:
\begin{equation}
    P = a \cdot \log_{10}(d) + b
    \label{eq:pathloss_model}
\end{equation}
where \( a \) and \( b \) are coefficients determined via linear regression, with \( P \) representing path loss and \( d \) the distance between transmitter and receiver.

This empirical fit is not intended as a novel propagation model but rather serves to summarize the measurement data, enabling direct comparison with existing theoretical and empirical models. The coefficients obtained are \( a = 24.8065 \) and \( b = 132.6223 \), with standard errors of 0.4862 and 0.1815 respectively. The fit yields an \( R^2 \) value of 0.7028, the $95\,\%$ confidence intervals are [23.8526, 25.7605] for \( a \) and [132.2662, 132.9783] for \( b \).

The comparison shows that theoretical models can deviate significantly from measured values, especially in complex urban areas where building density, construction materials, and topology affect signal propagation. For example, at $100~\mathrm{m}$, the Okumura-Hata model falls $14~\mathrm{dB}$ below our measurements, while at $5~\mathrm{km}$, it rises $7~\mathrm{dB}$ above, illustrating the impact of real-world conditions. Notably, the Beirut empirical model also diverges from our data, suggesting that even empirical models may not transfer well due to unique environmental factors.

Although our model provides meaningful results, it is affected by survivor bias, as it only includes successfully received packets. This bias leads to an underrepresentation of high-loss conditions, especially at longer distances, and limits the model's ability to capture extreme signal degradation. Consequently, the fitted model tends to underestimate path loss near the edge of the network's coverage. This limitation reinforces the importance of interpreting the fit within its intended scope: providing a compact representation of the observed data for comparative purposes, rather than predicting absolute path loss under all conditions.

These findings demonstrate the importance of broad data collection across varied environments for understanding signal behavior and improving prediction models. Field measurements allow models to be adapted to real-world conditions, increasing their reliability for practical deployments.

\subsection{RSSI Distribution Analysis}

\begin{figure}[H]
    \centerline{\includegraphics[width=0.495\textwidth]{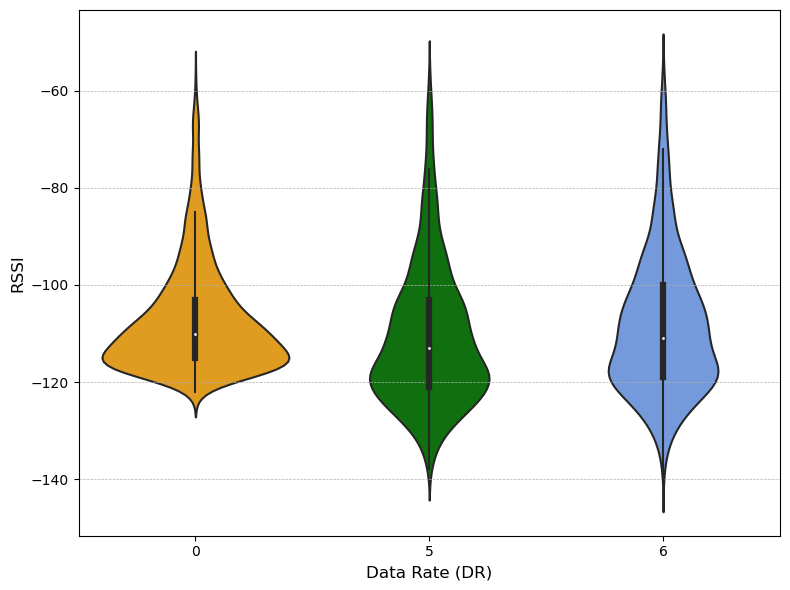}}
    \caption{Violin plot comparing the distribution of RSSI values for DR0, DR5, and DR6, illustrating variations in signal strength across the data rates.}
    \label{fig_rssi_violin}
\end{figure}

The violin plot in Fig.~\ref{fig_rssi_violin} provides insights into the distribution of Received Signal Strength Indicator (RSSI) values for DR0, DR5, and DR6. The minimum RSSI received is $-138~\mathrm{dBm}$ for DR5, $-137~\mathrm{dBm}$ for DR6, and $-120~\mathrm{dBm}$ for DR0. This suggests that LR-FHSS may achieve greater range, as indicated by its reported sensitivity.

\begin{table}[ht]
\centering
\caption{RSSI and Distance Statistics for Data Points Over $3~\mathrm{km}$ from the Gateway}
\setlength{\tabcolsep}{3pt} % Adjust the value as needed
\begin{tabular}{c|cccccc|c}
\hline
 & \multicolumn{6}{c|}{\textbf{RSSI (dBm)}} & \textbf{Distance (km)} \\
\textbf{DR} & \textbf{Mean} & \textbf{Std Dev} & \textbf{Min} & \textbf{Max} & \textbf{Median} & \textbf{Count} & \textbf{Mean} \\
\hline
0  & -115.44 & 3.09 & -120.0 & -103.0 & -116.0 & 48 & 3.31 \\
5  & -124.72 & 5.90 & -138.0 & -113.0 & -124.0 & 100 & 3.34 \\
6  & -122.65 & 5.82 & -137.0 & -112.0 & -121.5 & 46 & 3.31 \\
\hline
\end{tabular}
\label{table:summary_stats}
\end{table}

However, analyzing the average RSSI for data points over $3~\mathrm{km}$ from the gateway reveals a different perspective. Despite similar mean transmission distances, the average RSSI values differ significantly, as shown in Table~\ref{table:summary_stats}.

To eliminate environmental factors as a potential cause for the observed RSSI differences, we conducted localized tests at Fort Needham Memorial Park, which is elevated and approximately $3.3~\mathrm{km}$ from the gateway. The results, presented in Table~\ref{table:rssi_stats_dr}, consistently showed the same RSSI differences between LoRa and LR-FHSS, reinforcing our observations.

\begin{table}[ht]
\centering
\caption{Localized Measurements from Fort Needham Memorial Park}
\begin{tabular}{c|cccccc}
\hline
 & \multicolumn{6}{c}{\textbf{RSSI (dBm)}} \\
\textbf{DR} & \textbf{Mean} & \textbf{Std Dev} & \textbf{Min} & \textbf{Max} & \textbf{Median} & \textbf{Count} \\
\hline
0 & -114.27 & 3.67 & -118.0 & -103.0 & -115.5 & 22 \\
5 & -118.89 & 2.69 & -125.0 & -113.0 & -118.0 & 27 \\
6 & -119.39 & 3.68 & -127.0 & -112.0 & -118.5 & 18 \\
\hline
\end{tabular}
\label{table:rssi_stats_dr}
\end{table}

\begin{figure}[htbp]
    \centerline{\includegraphics[width=0.495\textwidth]{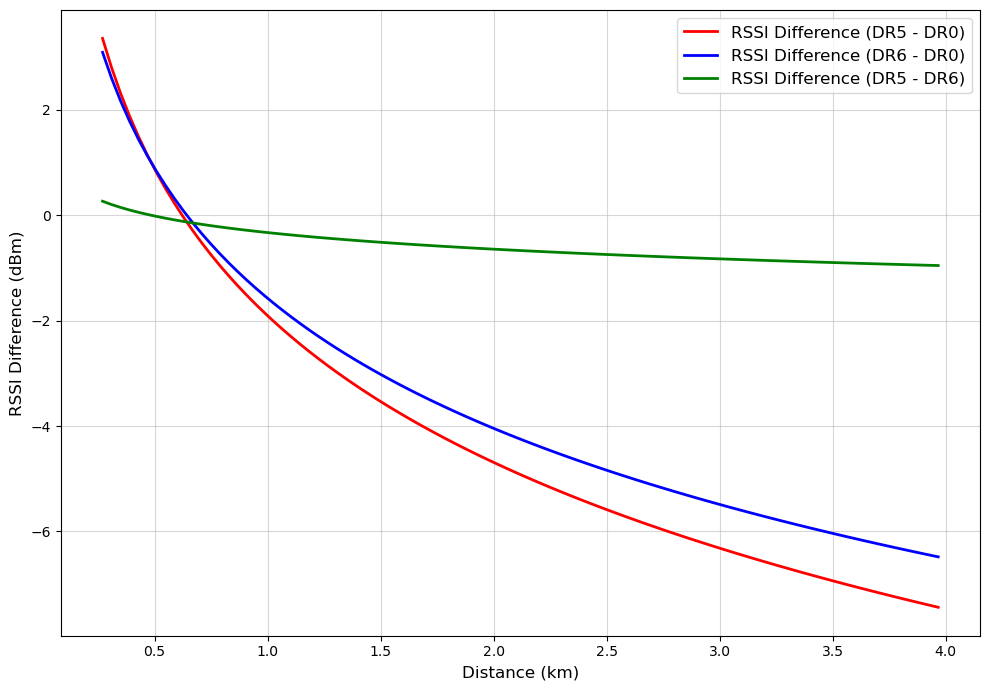}}
    \caption{RSSI differences between each data rate (DR0, DR5, DR6) as a function of distance from the gateway.}
    \label{fig_rssi_diff}
\end{figure}

We focused on data points far from the gateway, where the RSSI difference between LoRa and LR-FHSS was most pronounced. To understand how distance impacts this difference, we computed the RSSI difference as a function of distance using the logarithmic model introduced in equation~\eqref{eq:pathloss_model}. The differences were then plotted, as shown in Fig.~\ref{fig_rssi_diff}. It is evident that the difference between DR0 (LoRa) and DR5/DR6 (LR-FHSS) increases with distance.

It is important to note that while LR-FHSS appears to exhibit better RSSI values in our measurements, this does not necessarily imply superior receiver sensitivity. The observed differences must be interpreted cautiously, as RSSI is not a standardized metric, and its computation can vary depending on the modulation type. In our setup, both LoRa and LR-FHSS signals were received using the same Kerlink Wirnet iBTS Gateway. However, the gateway's internal RSSI estimation may differ depending on the modulation used, as modulation type influences hardware filtering, gain settings, and the specific demodulation paths employed within the receiver. This is particularly relevant given the substantial differences between LoRa's continuous chirps and LR-FHSS's rapidly hopping narrowband signals, which affect how power is accumulated and reported as RSSI. This is further illustrated by the similar packet reception rates observed between DR0 (LoRa) and DR6 (LR-FHSS), despite the notable differences in RSSI.

The RSSI difference between these modulations also relates to their transmission characteristics. LoRa transmits a continuous signal over a fixed bandwidth of $125~\mathrm{kHz}$, concentrating signal energy within this band for the entire packet duration. In contrast, LR-FHSS uses very narrow $488~\mathrm{Hz}$ signals that hop across a wide $1.523~\mathrm{MHz}$ band, spreading the same total energy over a much broader spectrum. This frequency hopping adds complexity to RSSI measurement, as the gateway must track and aggregate signal power across multiple rapidly changing sub-channels, which is a fundamentally different process than measuring a static LoRa signal.

The spectral power density reflects this difference:
\begin{itemize}
    \item For LoRa at $22~\mathrm{dBm}$ over $125~\mathrm{kHz}$: approximately $-29~\mathrm{dBm/Hz}$.
    \item For LR-FHSS at $22~\mathrm{dBm}$ over $1.523~\mathrm{MHz}$: approximately $-40~\mathrm{dBm/Hz}$.
\end{itemize}

This results in an approximately $11~\mathrm{dB}$ spectral power density difference, contributing to the observed RSSI differences between the two technologies. However, the increasing RSSI difference with distance, as shown in Fig.~\ref{fig_rssi_diff}, suggests that spectral power density alone cannot explain the trend. We attribute this behavior to the combined effects of the gateway's RSSI computation method and the distinct signal structures of LoRa and LR-FHSS.

\section{Conclusion}

This measurement campaign provides the first comprehensive real-world comparison of LR-FHSS and LoRa modulations within LoRaWAN networks in urban environments, specifically under the FCC-regulated US915 band. The results show that LR-FHSS significantly improved PRR, achieving up to a 20\,\% gain with DR5 in dense urban areas where interference is prevalent. However, PRR varied across data rates, with DR5 showing clear advantages due to its 1/3 coding rate, while DR6 did not consistently outperform LoRa's DR0.

The superior results with LR-FHSS (DR5) are attributed to its frequency hopping and stronger coding rate, which enhance resilience to interference and enable more reliable communication within LoRaWAN networks. These findings suggest that LR-FHSS can increase network efficiency in urban deployments and potentially reduce the number of gateways required for full city coverage. This reduction in infrastructure may lower deployment costs and simplify network management for LoRaWAN applications in constrained environments.

Despite these advantages, adoption of LR-FHSS in LoRaWAN remains limited by the availability of compatible devices and higher costs linked to the modulation's complexity. As the technology matures and more devices become available, these barriers are likely to diminish, making LR-FHSS a more accessible option for large-scale deployment.

The robustness and scalability of LR-FHSS also present opportunities for extending LoRaWAN connectivity via satellite communications~\cite{LoRaWAN_LR-FHSS_Webinar}. Its ability to operate over long distances and in areas with minimal infrastructure makes it well-suited for connecting remote or hard-to-reach regions. Future work will explore satellite integration, AI-driven network optimization, and the development of cost-effective hardware to fully leverage LR-FHSS in global LoRaWAN applications.

\section*{Acknowledgment}
The authors extend sincere gratitude to Olivier Seller from Semtech for his time and feedback on this article.

\bibliography{references}

\end{document}